\documentclass[aps,prb,twocolumn,showpacs,10pt,superscriptaddress]{revtex4-1}

\usepackage{amsmath}
\usepackage{bm}
\usepackage{graphicx}
\usepackage{amssymb}
\usepackage[caption=false]{subfig}
\usepackage{color}
\usepackage{verbatim}

\newcommand{\bep} {\boldsymbol{\epsilon}}
\newcommand{\bc} {\mathbf{c}}

\begin{document}

\title{Localisation and finite-size effects in graphene flakes}

\author{C. Gonz\'{a}lez-Santander}%
\affiliation{GISC, Departamento de F\'{i}sica de Materiales, Universidad
  Complutense, E-28040 Madrid, Spain}
%\thanks{E-mail: \email{cglezsantander@fis.ucm.es}}
\author{F. Dom\'{i}nguez-Adame}
\affiliation{GISC, Departamento de F\'{i}sica de Materiales, Universidad
  Complutense, E-28040 Madrid, Spain}
\author{M.\ Hilke}
\affiliation{Department  of Physics, McGill University, Montr\'{e}al (Qu\'{e}bec)
  H3A 2T8, Canada}
\author{R. A. R\"omer}
\affiliation{Department of Physics and Centre for Scientific Computing, University
  of Warwick, Coventry, CV4 7AL, UK}

%\date{$Revision: 1.26 $, compiled \today, \currenttime}

\begin{abstract}
We show that electron states in disordered graphene, with an onsite
potential that induces inter-valley scattering, are localised for all energies
at disorder as small as $1/6$ of the band width of clean graphene. We clarify
that, in order for this Anderson-type localisation to be manifested, graphene
flakes of size $\approx 200 \times 200\,$nm$^{2}$ or larger are needed. For smaller
samples, due to the surprisingly large extent of the electronic wave functions,
a regime of apparently extended (or even critical) states is identified. Our
results complement earlier studies of macroscopically large samples and can
explain the divergence of results for finite-size graphene flakes.
\end{abstract}

\pacs{
73.22.Pr %Electronic structure of graphene
73.20.Fz %Weak or Anderson localisation
}

\maketitle

%%%%%%%%%%%%%%%%%%%%%%%%%%%%%%%%%%%%%%%%%%%%%%%%%%%%%%%%%%%%%%%%%%%%%%%%%%
\section{Introduction}
%%%%%%%%%%%%%%%%%%%%%%%%%%%%%%%%%%%%%%%%%%%%%%%%%%%%%%%%%%%%%%%%%%%%%%%%%%

In two-dimensional (2D) quantum systems, uncorrelated potential disorder has
been shown to lead to complete (Anderson) localisation of single particle states
\cite{And58,LeeR85,KraM93,EveM08}. This statement has been supported by a wealth
of experimental, numerical and theoretical results, including the celebrated
scaling hypothesis~\cite{AbrALR79} and seminal works based on the nonlinear
$\sigma$ model~\cite{EveM08,BelK94}. States in a 2D system are marginally
localised even for small disorder and $d=2$ is the lower critical dimension of
the Anderson transition in time-reversal invariant systems. However, while this statement is true in general, it has also been shown that the situation is more complex when correlations in the
disorder~\cite{BelDHT99,IzrK99} or many-body interactions~\cite{BelK94,PunF05}
have to be taken into account. Even without these additional factors, the 2D
situation remains challenging since the extent of the localised states for weak
disorder can become much larger than the available system sizes, which
might lead to results of a feigned extended behaviour.

In graphene, as prototypical 2D material~\cite{NovGMJ04,Gei09}, one naturally
expects disorder to lead to localisation as well. However, due to its linear
dispersion relation around the Dirac point at energy $E=0$ and non-zero momentum, the
resulting absence of backscattering in clean samples~\cite{AndNS98}, might lead to 
a somewhat unusual behaviour.
The localisation properties of graphene in the vicinity of the Dirac point have
been studied intensively. It was found that strong disorder leads to
localisation at $E=0$~\cite{SchO92,XioX07}, while disorder that does not
lead to inter-valley mixing does not~\cite{BarTBB07,NomKR07}. The direction of
transport along graphene~\cite{LheBNR08} and graphene
nanoribbons~\cite{SchSF09,GunW10} was shown to modify the quantitative strengths
of the localisation effects.
On the other hand, many, mainly numerical, results have indicated the existence
of localisation that is unusually weak at
$E=0$~\cite{AmiJS09,AmaE09,SonSF11,BarN12,AmaKKE13} or close to
$E=0$~\cite{Hil10}. Some results supporting mobility
edges~\cite{AmiJS09,SonSF11}, critical states~\cite{AmaE09,BarN12,AmaKKE13}
and a metallic-like to insulating transition~\cite{Hil10}
have been put forward.
Recent discussions of results at $E=0$~\cite{SchSF10,AmiJS10} or for strong
disorder at $E\gtrsim 0$~\cite{LeeGMD13} indicate complete localisation for
disorder with inter-valley mixing, in agreement with the earlier studies
\cite{XioX07,BarTBB07,NomKR07} and a true metal-insulator transition has only
been observed in hydrogenated graphene~\cite{BanC10,SchF12}.

\begin{figure}[tb]
\includegraphics[width=0.95\columnwidth,clip]{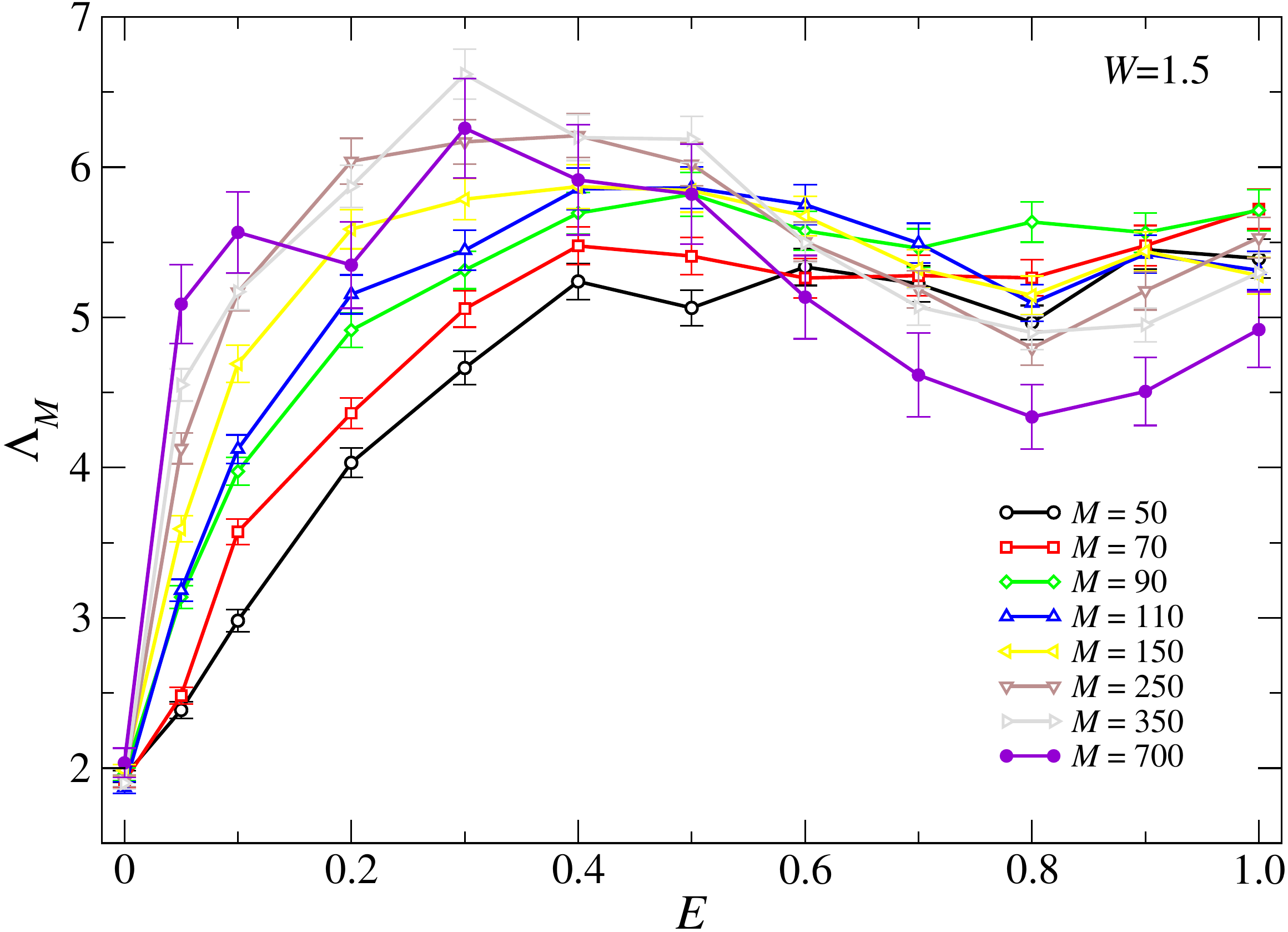}\\[-23ex]
\hspace*{10ex}
\includegraphics[width=0.33\columnwidth,clip]{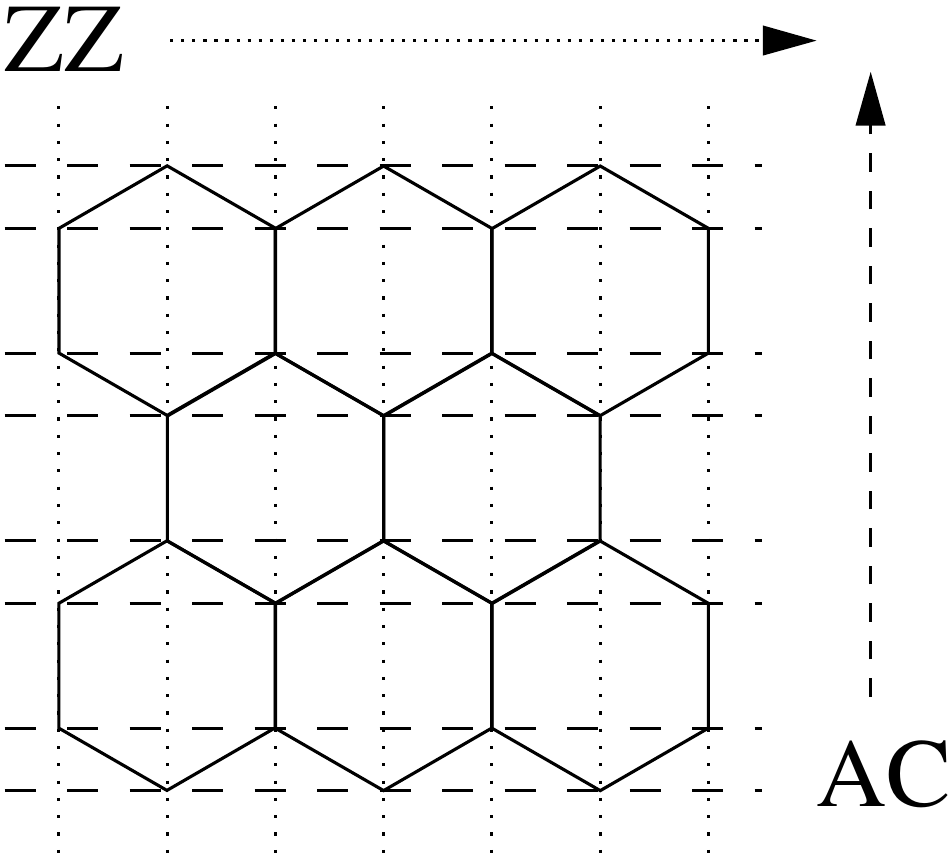}\\[1ex]
\caption{(Color online)  Reduced localisation length $\Lambda_M=\lambda_M/M$ as
a function of energy for the ZZ graphene lattice at disorder $W=1.5$ and
sizes ranging from $M \times L=50 \times 50$ to $M \times L=700 \times 700$. The
error bars indicate the error of the mean from averaging over $500$ samples,
except for $M \times L=700 \times 700$ where the average is over $100$ samples.
The lines are guide to the eye only.
Inset: schematic of the AC and ZZ edge transport directions (arrows) while the
dashed (dotted) lines indicate AC (ZZ) transverse layers, respectively (here $M\times L=4\times 7$ in ZZ).
}
\label{fig-ENE-01}
\label{fig-lambdaM_E_W1.5}
\label{fig-lattice}
\end{figure}

Nevertheless, these studies still leave the regime of small energies that are close to
but away from $E=0$, for weak but inter-valley mixing onsite disorder
unresolved, where Ref.~\onlinecite{Hil10} (see their fig.~2) found evidence for a
transition-like behaviour. In fig.~\ref{fig-lambdaM_E_W1.5} we show this
behaviour for 2D graphene flakes with $700^2$ lattice sites. Clearly, increasing the size $M^2$ of the graphene samples leads to
increasing localisation lengths around
$E\approx 0.25$, with energy in units of the hopping energy between
carbon atoms, while around $E=0.9$ the trend seems to have
reversed.
In this paper, we will show that fig.~\ref{fig-lambdaM_E_W1.5} does not
indicate the existence of a transition to delocalised states. Rather, we find
that the finite-size trend reverses towards localised behaviour upon further increasing
the system size. However, we will need to go to very large system sizes of the
order of $2.25\times 10^6$ lattice sites to show this. For smaller system sizes
from about  $360,000$ to about
$10^6$ ~\cite{AmiJS09,AmaE09,SonSF11,BarN12,AmaKKE13}, scaling results indicate
roughly a system size independence of $\Lambda_M$.  Hence our  results explain
why there is such a diversity of results for the localisation  properties of
graphene at and close to $E=0$, i.e.\ we find that one needs very  large system
sizes, larger than $2\times 10^6$ lattice sites, to reach the asymptotic regime.

%%%%%%%%%%%%%%%%%%%%%%%%%%%%%%%%%%%%%%%%%%%%%%%%%%%%%%%%%%%%%%%%%%%%%%%%%%
\section{Numerical approach}
%%%%%%%%%%%%%%%%%%%%%%%%%%%%%%%%%%%%%%%%%%%%%%%%%%%%%%%%%%%%%%%%%%%%%%%%%%

Our calculation is based on the standard 2D single-particle Hamiltonian
\begin{equation}
\mathcal{H}=\sum_{l=1}^{L}  \bc^\dagger_l \bep_l \bc_l - \sum_{l=1}^{L-1}
(\bc^\dagger_l \mathbf{t}_l \bc_{l+1} +\bc^\dagger_{l+1} \mathbf{t}_l\bc_l),
\label{eq-ham}
\end{equation}
on a lattice with $L \times M$ sites. Here, $\bep_l$ denotes the $M\times M$
Hamiltonian matrix acting in the (transverse) $m$ direction for each vertical
arm at (longitudinal) position $l$~\cite{RodCR12} and $\bc_l^\dagger \equiv
\left( c_{l,1}^\dagger, c_{l,2}^\dagger, \ldots, c_{l,M}^\dagger \right)$, with
$c_{l,m}$ ($c^\dagger_{l,m}$) the usual annihilation (creation) operators of a
tight-binding orbital at the site $\{l,m\}$. The diagonal elements for each
$\bep_l$ correspond to random onsite potentials $\epsilon_{l,m}\in [-W/2,W/2]$,
$m=1,  \ldots, M$, which are uniformly distributed and $W$ determines the disorder strength. The off-diagonal  elements
model the hopping in transverse direction while $\mathbf{t}_l\equiv
t\,\bm{C}_l$ is the hopping along the $l$ direction  with $\bm{C}_l$ denoting
the connectivity matrix between layers $l$ and
$l+1$~\cite{EilFR08,SchO92,LeeGMD13}. All energies are measured in units of the
hopping energy, $t$.

The electronic problem defined by the Schr\"{o}dinger equation
$\mathcal{H}\psi=E\psi$ for the Hamiltonian \eqref{eq-ham} can be studied
conveniently by the transfer-matrix method (TMM)~\cite{KraM93,LeeGMD13}.
However, since we are not interested in the quasi-1D problem of graphene
nanoribbons with $L\gg M$~\cite{SchSF09,GunW10}, we need to modify the TMM to
allow the treatment of 2D $M \times M$ graphene samples.~\cite{foot1}
%
% \footnote{A ``square'' $M\times M$ sample for, e.g.\ ZZ graphene physically
% corresponds to a rectangle  with length to width ratio of $\sqrt{3}M/(3M-2)\sim
% 0.58$. For AC graphene, the ratio is $(3M-2)/\sqrt{3}M$.}
%
This has implications for the convergence of standard TMM calculations since we
can no longer use the self-averaging property normally used for $L \rightarrow
\infty$.
Our modification involves the definition of forward and backward transfer matrix
multiplications~\cite{FraMPW95,NdaRS04}. The method also yields the inverse
localisation length $1/\lambda_M(E,W)$, but only for a single $M\times M$
graphene sample. Afterwards, the $1/\lambda$ values need to be averaged for many
$M\times M$ disorder configurations with the same parameters $M$, $E$ and $W$.

The TMM must be adapted to handle the hexagonal structure of the graphene
lattice~\cite{SchO92,EilFR08} by suitably chosen  $\bep_l$ and $\bm{C}_l$
matrices. We distinguish between transport directions  parallel to
armchair~(AC) and zig-zag~(ZZ) edges. Our approach is similar to
Ref.~\onlinecite{LeeGMD13} and for more details see Ref.~\onlinecite{Gon13}.
A pictorial representation is shown in the inset of fig.~\ref{fig-lattice} for
AC and ZZ graphene.~\cite{foot2}
%
% \footnote{The choice of  layers in fig.~\ref{fig-lattice} leads to equal spacing
% for ZZ graphene  with inter-layer distance $0.142$nm $\times \cos \pi/6 =
% 0.123$nm. For AC  graphene the inter-layer spacing alternates between $0.142$nm
% and $0.142$nm  $\times \sin \pi/6=0.071$nm.  Similar considerations apply in the
% transverse direction. We will not  attempt to rescale these length scales here.}
%
We chose hard wall boundary conditions for all results presented here. In order
to have the same number of atoms for both ZZ  and AC edges, the width  of the AC
sample should be chosen as $M_{\text{AC}}=L_{\text{ZZ}}/2$ and the  length as
$L_{\text{AC}}= 2 M_{\text{ZZ}}$. In this way we ensure that we are  studying
the same sample but in both directions of transport.

The scaling hypothesis for finite-sized systems implies
$\Lambda_M(E,W)\equiv\lambda_M(E,W)/M = f(\xi(E,W)/M)$ for a suitably chosen
scaling parameter $\xi(E,W)$~\cite{LeeR85}. For strong disorder, $\lambda_M
\propto  \xi$~\cite{KraM93}. The $\lambda_M$ data can be rescaled numerically by
a  least-squares fitting procedure to obtain the scaling function
$f$~\cite{MacK81,KraM93}. In the case  of the 2D Anderson model on a square
lattice, this function has a single  finite-size scaling (FSS) branch with
decreasing $\Lambda_M$ for increasing $M$ --- indicating the  localised regime.
For the 3D Anderson model, the same procedure leads to two  branches, the first
one denoting the localised regime and the second one  indicating the extended
regime with increasing $\Lambda_M$ values as $M$  increases. This two-branch
behaviour is the signature of the transition from  localised to extended
states~\cite{KraM93}.
Alternatively, we can try to assume an analytical form for $f$ and test whether
this form fits the data with the required accuracy~\cite{SleO99a,RodVSR11}.
Assuming e.g.\ the power-law behaviour $f\propto |1-E/E_\text{c}| L^{1/v}$ of
the 3D Anderson transition, then this approach allows not only to construct
$f$, but also determines the critical exponent $\nu$ and the energy
$E_\text{c}$ (or disorder $W_\text{c}$) at which the transition
occurs~\cite{SleO99a,RodVSR10}. We will use both FSS approaches below.

%%%%%%%%%%%%%%%%%%%%%%%%%%%%%%%%%%%%%%%%%%%%%%%%%%%%%%%%%%%%%%%%%%%%%%%%%%
\section{Results}
%%%%%%%%%%%%%%%%%%%%%%%%%%%%%%%%%%%%%%%%%%%%%%%%%%%%%%%%%%%%%%%%%%%%%%%%%%

\begin{figure*}[tb]
\centering
\includegraphics[width=0.565\columnwidth,clip]{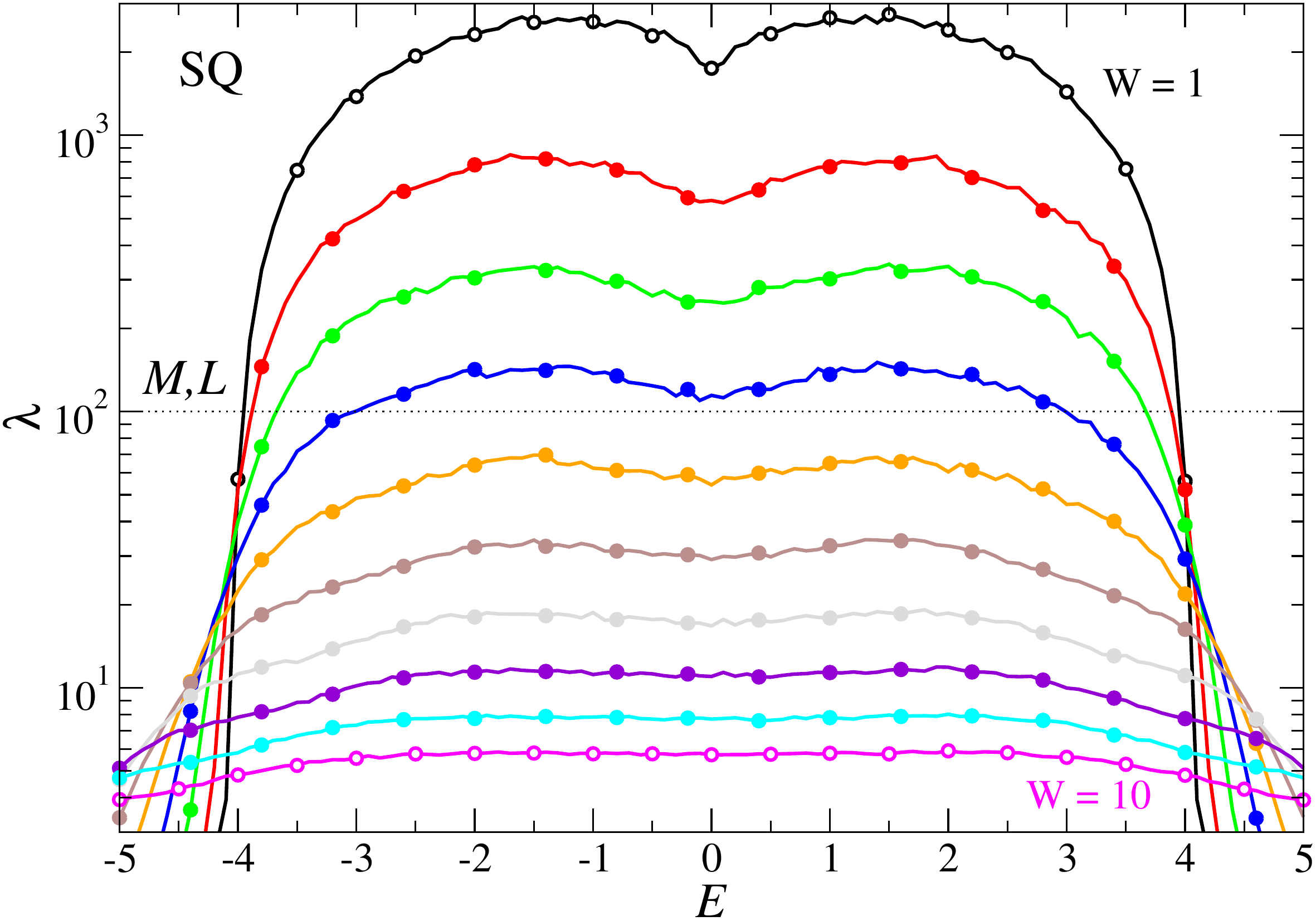}
\includegraphics[width=0.52\columnwidth,clip]{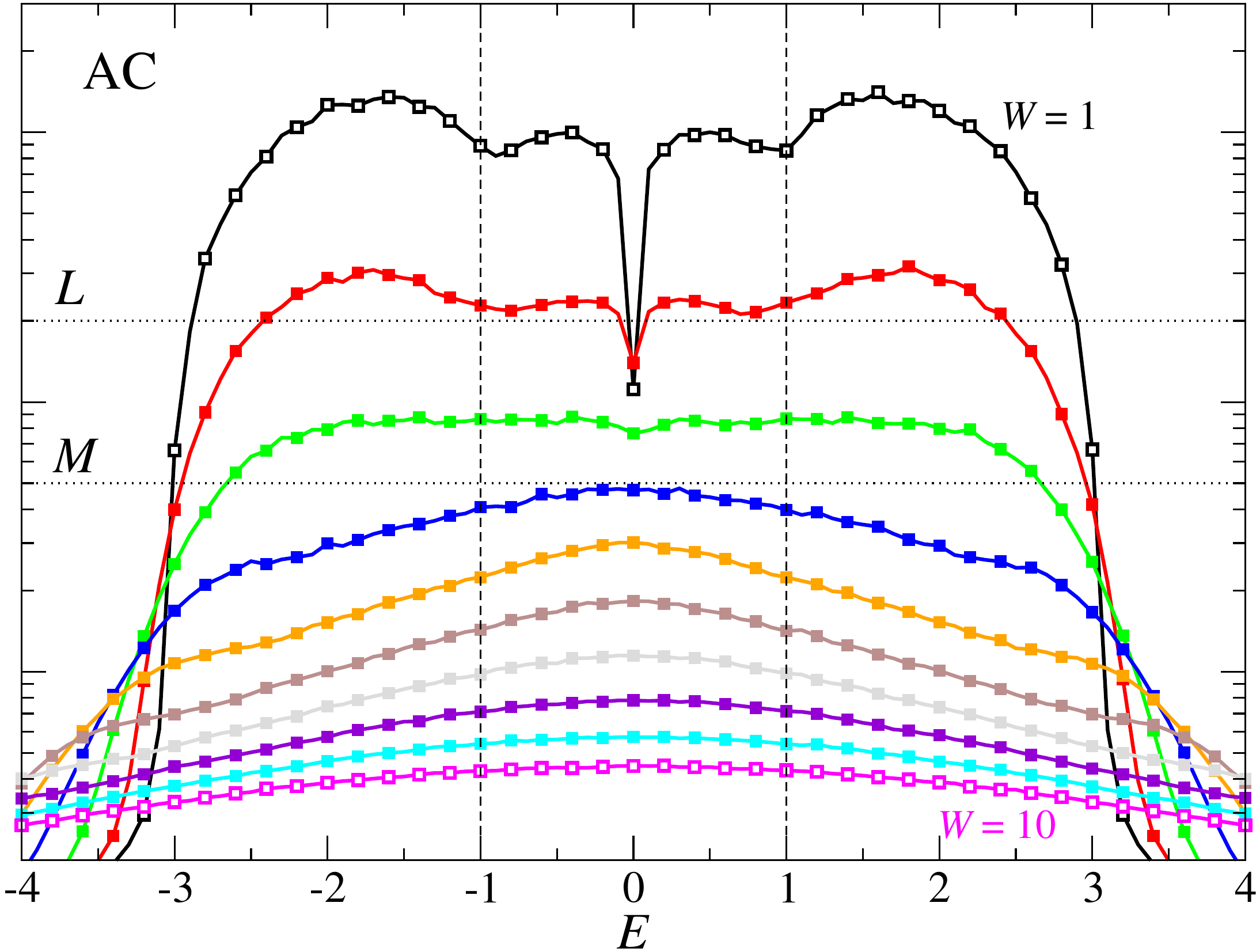}
\includegraphics[width=0.545\columnwidth,clip]{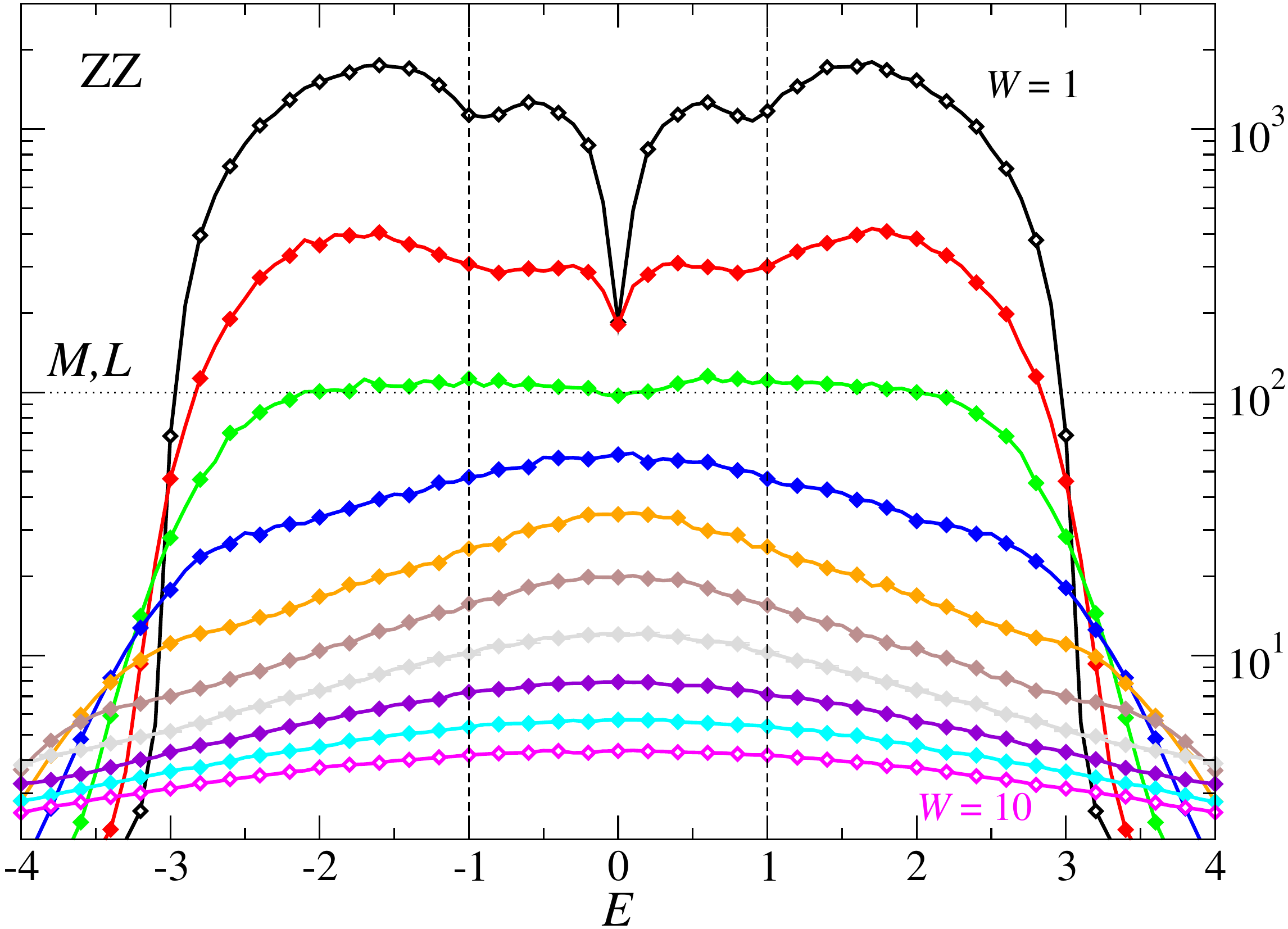}
\caption{(Color online) Average localisation length as a function of energy for
a square lattice (left panel), AC (central panel) and ZZ (right panel) graphene
lattices for systems with $10^4$ lattices sites and different values of disorder $W=1, 2,
\ldots, 10$. Lines connecting the data values are guides to the eyes only. For
clarity, we only indicate the labels for $W=1$ and $W=10$. The error bars are
within the symbol sizes. The $M$ and $L$ values are indicated by horizontal
dashed lines. The vertical lines at $E=\pm 1$ for AC and ZZ graphene mark the
position of the van-Hove singularities in the density of states of clean
graphene.}
\label{fig-ENE-SQZZAC}
\end{figure*}
%%%%%%%%%%%%%%%%%%%%%%%%%%%%%%%%%%%%%%%%%%%%%%%%%%%%%%%%%%%%%%%%%%%%%%%%%%

In fig.~\ref{fig-ENE-SQZZAC} we show the variation of the disorder-averaged
localisation length $\lambda_M(E)$ for different values of disorder $W$. The lattices
correspond to square lattice, AC and ZZ graphene. In each case, the system sizes where
chosen such that $M\times L = 10^4$ lattice sites, corresponding to
$M=100$ and $L=100$ for the square lattice and the ZZ graphene, but $M=50$ and
$L=200$ for the AC graphene lattice.
We first note that at weak disorder ($W\sim 1$) the half of the bandwidth reflects the
number of nearest neighbours and hence tends to $4$ for the square lattice and
tends to $3$ for AC and ZZ graphene~\cite{Ger31}. Furthermore, there is the usual
approximate symmetry between positive and negative energies.
When the strength of the disorder increases the $\lambda$ values decrease for
all lattices as the wavefunctions become more localised. For very strong
disorder, the localisation lengths are much smaller than the system sizes $M$
and $L$ and the states are exponentially localised with $\lambda$ representing
the decay length. On the other hand, for weaker disorder, the localisation
lengths are comparable or larger than the system sizes, and we can no longer
assume that the exponential decay implicit in the use of $\lambda$ is still
justified. Then $\lambda$ is simply a convenient measure of the spatial extent
of the wave functions, but not necessarily linearly related to a localisation
length. Still, a larger such extend will imply larger $\lambda$ values.
With this in mind, we see in fig.~\ref{fig-ENE-SQZZAC} that, for $W \lesssim 4$,
the localisation lengths increase rapidly as we decrease $W$ for the square
lattice. However, for the case of AC and ZZ graphene lattices, we observe that
in the vicinity of $E=0$, the $\lambda$ values again decrease, leading to values
of $\lambda_M(E\approx 0)$ which seem very similar for $W=1$ and $2$. Clearly,
the drop in $\lambda_M$ in the graphene lattices at $E=0$ is a  signature of the
Dirac point with reduced density of states~\cite{Wal47,GriRS98}.

In standard quasi-1D TMM, an increasing value of $\Lambda_M$ for weak disorder
as $M\rightarrow\infty$ signals the start of the extended  regime. Even with
$\Lambda_M>1$, $\lambda_M$ can still be interpreted as a localisation length
since we have $L\gg M$ and the localisation in the $l$ direction is well-defined.
As discussed before, the situation might be different for our modified TMM.
Nevertheless, we already see from fig.~\ref{fig-ENE-SQZZAC} that for energies
$|E|\gtrsim 1$, the $\lambda$ values for the square lattice and AC/ZZ graphene
behave similarly. If  any new, graphene-specific,  finite-size behaviour
can be expected, it should be around $E\approx 0$.
Therefore we have studied in fig.~\ref{fig-ENE-01} the finite size behaviour of
$\Lambda_M$ in ZZ graphene for energies $0\leq E \leq 1$ at weak disorder
$W=1.5$ when $\Lambda_M \geq 1$. As one can see from this figure, for energies
larger then $E=0.9$, increasing $M$ (and $L$) leads to a decrease of
$\Lambda_M$, the traditional signature of localisation. However, for energies
$E\lesssim 0.6$, increasing $M$ gives {\em increasing} $\Lambda_M$ values. Such
a behaviour for $M\rightarrow\infty$ would indicate extended  states. Quite
similar findings have been reported previously in the same energy range for
smaller systems up to $M=252$~\cite{Hil10}.

Clearly, the existence of extended states in the vicinity of the Dirac point in
weakly disordered (but with inter-valley mixing) graphene would be surprising.
However, let us already note several suspicious observations, namely (i)~there
is no clear crossing point, rather a series of not well-defined crossing points
in the region $E \in [0.7,0.9]$. Furthermore,~(ii) increasing the system size
does not lead to a clearer crossing, and we can also not identify a simple,
monotonic in $M$ (irrelevant) shift of such a crossing point. Let us also
emphasize that system widths of $M=700$ as used in fig.~\ref{fig-ENE-SQZZAC} are
already reasonably large for TMM~\cite{SleO99a}.
If there truly was a metal-insulator transition in the indicated energy range,
then we would expect to see good quality FSS. On the other
hand, if the behaviour of fig.~\ref{fig-ENE-01} was simply due to finite-size effects, then we should see the increase in
$\Lambda_M$ vanish for large enough $M$. Since the increase seems largest
at energy $E=0.25$, we shall study this energy in detail for a square lattice as well as
AC/ZZ graphene.

%%%%%%%%%%%%%%%%%%%%%%%%%%%%%%%%%%%%%%%%%%%%%%%%%%%%%%%%%%%%%%%%%%%%%%%%%%
\begin{figure*}[tb]
\includegraphics[width=0.95\columnwidth,clip]{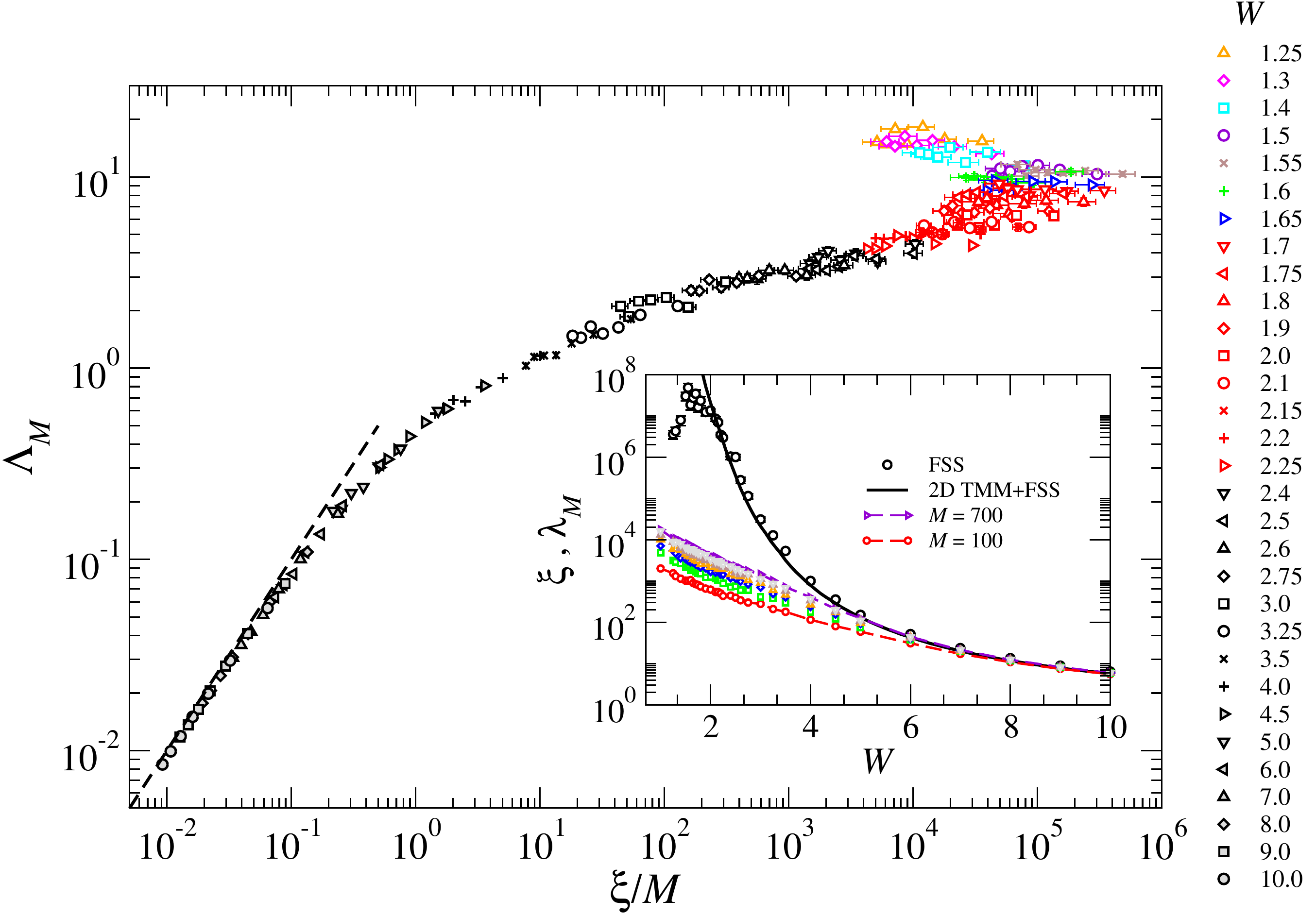}
\includegraphics[width=0.95\columnwidth,clip]{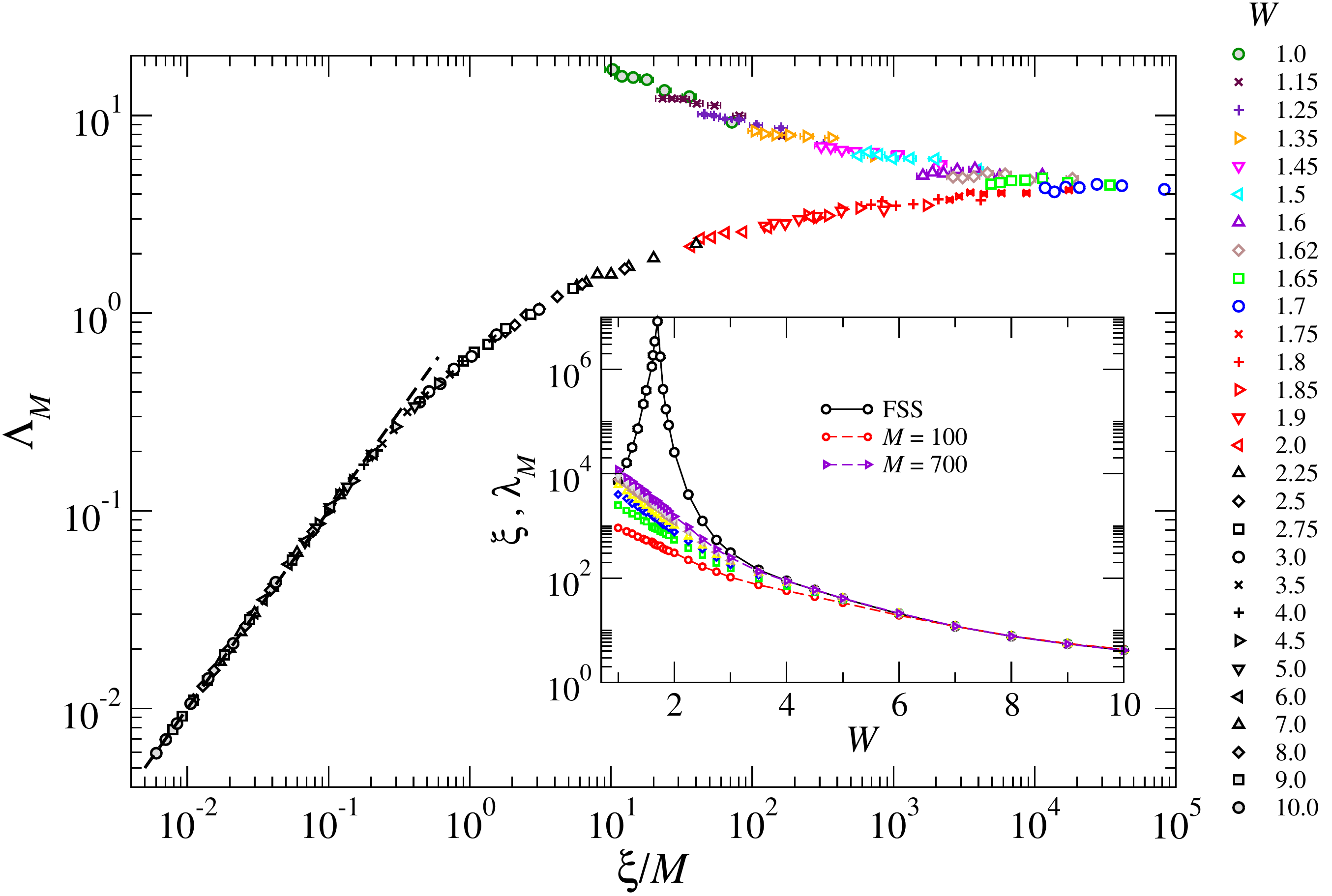}
\caption{Reduced localisation length as a function of reduced scaling parameter
for square lattices (left plot) and ZZ graphene  lattice (right plot) at
$E=0.25$. The disorder values used are $W\in [1, 10]$ as indicated in the
legends. The dashed line in each plot indicates the expected relation
$\lambda_M(W) \propto \xi(W)$ for large $W$. The error bars are only shown when
larger than the symbol sizes and have been generated by resampling the FSS according
to the accuracy of each $\Lambda_M$ value~\cite{Gon13}.
The inset in all cases shows $\xi(W)$ scaled to coincide with $\lambda_M$ values
for large $W$. The solid line in the inset of the left plot corresponds to
$\xi(W)$ obtained after FSS of standard TMM localisation lengths in quasi-1D
square lattices~\cite{LeaRS99a}. The other lines in the inset are guides to the
eye only.}
\label{fig-SQZZ-rLLvsKsis}
\end{figure*}
%%%%%%%%%%%%%%%%%%%%%%%%%%%%%%%%%%%%%%%%%%%%%%%%%%%%%%%%%%%%%%%%%%%%%%%%%%

In fig.~\ref{fig-SQZZ-rLLvsKsis} we show FSS results for $\Lambda_M$ in square
lattices and ZZ graphene with $M$ and $L$ values chosen such that the number of
sites $M\times L$ ranges from from $100^2$ to $700^2$.  For strong disorder, we
have $\Lambda_M\propto 1/M$ as expected since states  are highly localised and
$\lambda_M$ is constant for $M\gg\lambda_M$ as indicated.
Decreasing the disorder --- or, equivalently, decreasing $M$ --- leads to
deviations from the simple $1/M$ behaviour and indicates that $\xi(W)$ starts to
increase. In the standard quasi-1D square lattice TMM, this leads to an evermore
flat behaviour for $\Lambda_M(W)$ as $W\rightarrow 0$. We indeed  observe this
behavior for $E=0$ in square lattices, AC and ZZ graphene (not
shown)~\cite{Gon13}. For smaller  disorder, $W \lesssim 2$, we find the
reconstruction of a well-defined FSS curve becomes numerically difficult.
Nevertheless, the estimated scaling parameter $\xi(W)$ agrees very well with a
previous high-precision FSS from a quasi-1D TMM  \cite{LeaRS99a}. Furthermore,
the $\xi(W)$ behavior for squares and ZZ graphene shows a single branch only,
consistent with complete localisation.

The situation is rather different for $E=0.25$ as shown in
fig.~\ref{fig-SQZZ-rLLvsKsis}. We see that FSS  gives rise to  localised
branches as well as the beginnings of what look like extended  branches. Here it
is intriguing to see that even for a square lattice, for the range of available
system sizes and disorder --- determined by the longest TMM runs available to
us --- we find an apparent transition-like behavior. Obviously, this would be in
disagreement with the scaling theory and of course also to the body of numerical
results based, among others, on quasi-1D TMM~\cite{KraM93,EveM08}.
Similarly, we observe transition-like behavior also for ZZ graphene at $E=0.25$.
As in the square lattice case, the onset of the extended branch is around
$W\lesssim 2$. We have found similar results also for AC graphene.~\cite{foot3}
%
% \footnote{Previous studies have demonstrated that the finite width of AC flakes
% determines its metallic or insulating behaviour at $E=0$ besides the
% disorder~\cite{NakFD96}. In the case that $2M_{\text{AC}}=3n-1$ with $n\in
% \mathbb{Z}$, the system is metallic. At weak disorder, we observe this effect
% in our calculation through a large increase of $\lambda_M$ for such
% $M_{\text{AC}}$ that satisfies the above condition. At $E=0.25$ we do not see
% these finite width effects. Here we always choose $2M_{\text{AC}}\neq 3n-1$ in
% order to avoid the edge-state metallicity at these special sizes for
% $W=0$~\cite{NakFD96}.}

We have also tried to apply FSS assuming the expansions of the power-law
behaviour~\cite{SleO99a,RodVSR10}. However, we never find  an acceptable fit to
the data, although we vary not only the expansion  coefficients, but also the
initial values used in the non-linear fits for  $W_c$,  $\nu$, etc. Upon closer
inspection, we find that most such attempts to fit the  data lead to $W_c \sim
0$ and large values of $\nu > 5$. But even with these  large $\nu$ values,  the
$\Lambda_M$ values rise much faster for small disorder. This suggests that  the
true behaviour is not a power-law but rather an exponential as in the well-known
square lattice~\cite{MacK83}.

The FSS results of fig.~\ref{fig-SQZZ-rLLvsKsis} for $E=0.25$ and $W\lesssim 2$
do not show a very clear formation of extended branches, particularly for the
square case. In order to test the stability of these branches in FSS, we would
need even larger system sizes for {\em all} disorder $W\lesssim 2$. This is,
however, numerically prohibitive.~\cite{foot4}
%
% \footnote{The calculation of a single $M\times L= 700 \times 700$ sample for ZZ
% graphene can  take more than $6$ hours at small $W\lesssim 2$ one a single
% processor core.  The computing time increases to about a week for some
% $1500\times 1500$ samples.}
%
Thus we have chosen to restrict ourselves to two disorder strengths, $W=1$ and $1.25$ for
$E=0.25$. Even with this restriction, a considerable number of runs for $M>900$
do not finish within our chosen maximum time limit of about one week. Such
$\lambda_M$ values have therefore a relative error $\epsilon_n$, with $n$
denoting the sample, larger than the target of  $\epsilon_0\equiv 5\times
10^{-5}$. Hence we weigh such results less when computing  an average. With
(i)~$w_n=1/\epsilon_n^2$ or (ii)~$w_n = \text{max}(1,\epsilon_0/\epsilon_n)$, we
define the averaged Lyapunov exponents as $\overline{\gamma}_M=
\sum_{n}w_n\gamma_n/\sum_n w_n$ with weighted standard-deviations $\sqrt{\sum_n
w_n(\gamma_n-\overline{\gamma}_M)^2/\sum_n w_n}$. In case (ii),
samples, which have converged better than the target, are given less weight in order to test the
robustness of our results.

\begin{figure}[tb]
\includegraphics[width=\columnwidth,clip]{%
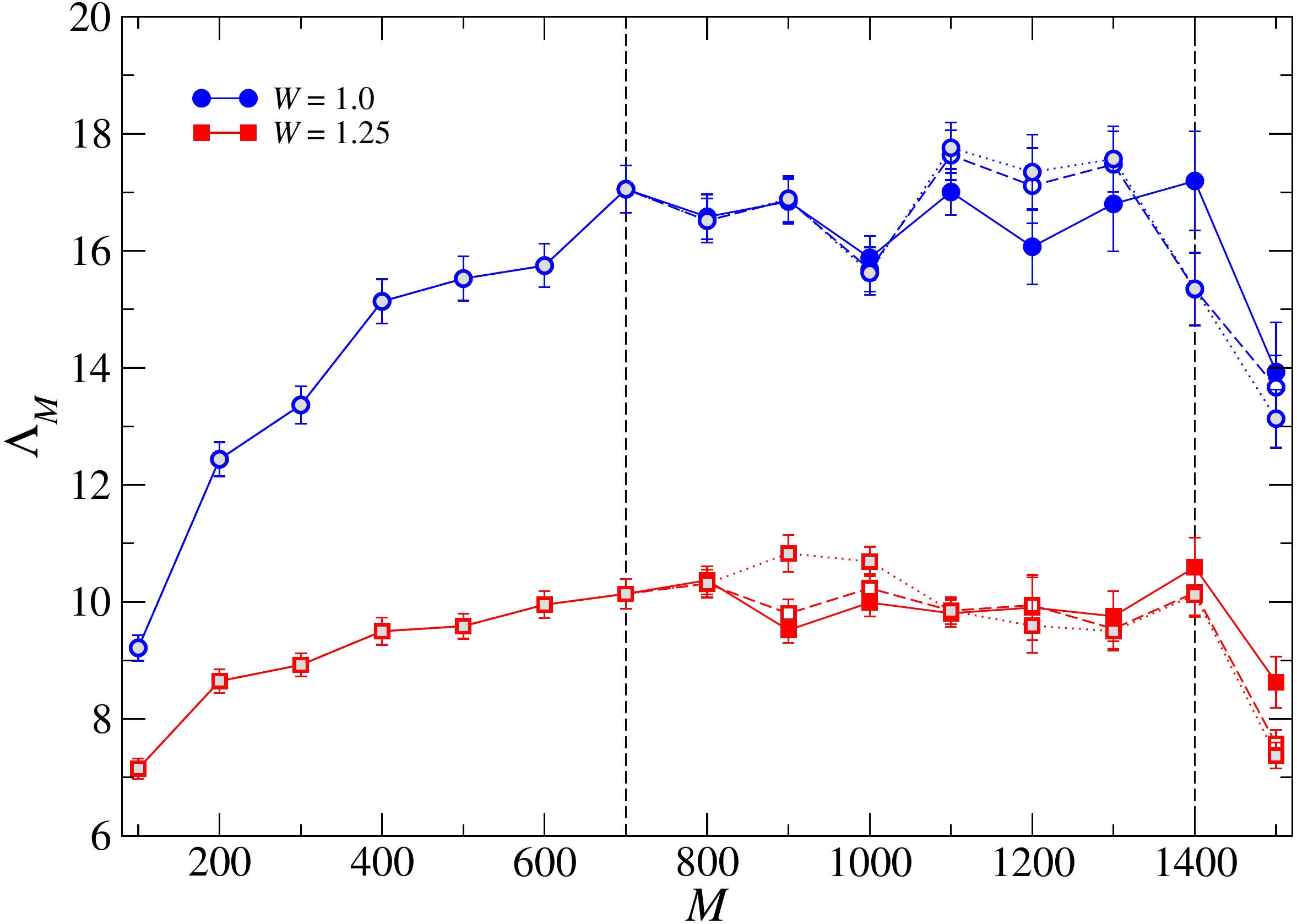}
\caption{Reduced localisation length as a function of $M$ for ZZ graphene at
$E=0.25$ with $W=1.0$ and $1.25$. The error bars indicate the error of the mean. The
mean itself has been computed using (solid symbols) the standard arithmetic
average $\sum_\mathrm{samples} \lambda_M^{-1}$ as well as (open/greyed symbols)
two weighted means as explained in the text. The data lines are guides to the
eye only. The vertical lines indicate regions of different size dependence for
$\Lambda_M$.}
\label{fig-ZZ-rLLvsM-E0-E025}
\end{figure}

We show the resulting system size dependence of $\Lambda_M$ values up to
$M=1500$ in  fig.~\ref{fig-ZZ-rLLvsM-E0-E025}.
%%%%%%%%%%%%%%%%%%%%%%%%%%%%%%%%%%%%%%%%%%%%%%%%%%%%%%%%%%%%%%%%%%%%%%%%%%
%%%%%%%%%%%%%%%%%%%%%%%%%%%%%%%%%%%%%%%%%%%%%%%%%%%%%%%%%%%%%%%%%%%%%%%%%%
%
We see that up to $M=700$, the $\Lambda_M$ values increase with increasing $M$,
as for extended states. From $M=800$ onwards, there is a regime in which we see
little or no dependence on $M$ within the fluctuations of the data. Such
behaviour, if it were to continue for $M\rightarrow\infty$, would be indicative
of critical states. Finally, at $M=1500$, we find a drop in $\Lambda_M$. The
drop is present both in the unweighted mean as well as, and even stronger, in
the weighted means.
This indicates that the observed increase in $\Lambda_M$ with increasing $M$ up
to $M=1400$ is simply a finite-size effect. Going to larger system sizes
recovers the expected behaviour for localised states with decreasing $\Lambda_M$
for increasing $M$. The hypothetical ``extended'' FSS curves in
fig.~\ref{fig-SQZZ-rLLvsKsis} should hence be interpreted as an intermediate
regime in which the localisation lengths become very large. Indeed, with $\Lambda_M
\approx 10$, this is beyond what has been observed in most previous TMM
studies.

A qualitative argument can be put forward to motivate our results. Without
disorder the density of states (DOS) at $E=0$ for a square lattice diverges
whereas it is zero for graphene (ZZ or AC). Upon increasing the disorder, the
DOS for the square lattice decreases as well as the localisation length. For
graphene, the same happens for the van-Hove singularities at $E=\pm
1$~\cite{GriRS98}. On the other hand, at $E=0$ the DOS {\em
increases}~\cite{GriRS98}, which is well-known to correlate with large
localisation lengths. The crossover between these two regimes should be expected
around $E \approx 0.5$ which is similar to what we observe. For larger disorder
$W \gtrsim 2$ or, equivalently, larger system sizes $M\gtrsim 1400$, we recover
the expected localised regime.

%%%%%%%%%%%%%%%%%%%%%%%%%%%%%%%%%%%%%%%%%%%%%%%%%%%%%%%%%%%%%%%%%%%%%%%%%%
%\subsection{Wave functions}

%%%%%%%%%%%%%%%%%%%%%%%%%%%%%%%%%%%%%%%%%%%%%%%%%%%%%%%%%%%%%%%%%%%%%%%%%%
\begin{figure}[tb]
\centering
\subfloat[]{\label{fig-WF-ZZ-W05}\includegraphics[width=0.25\columnwidth,clip]{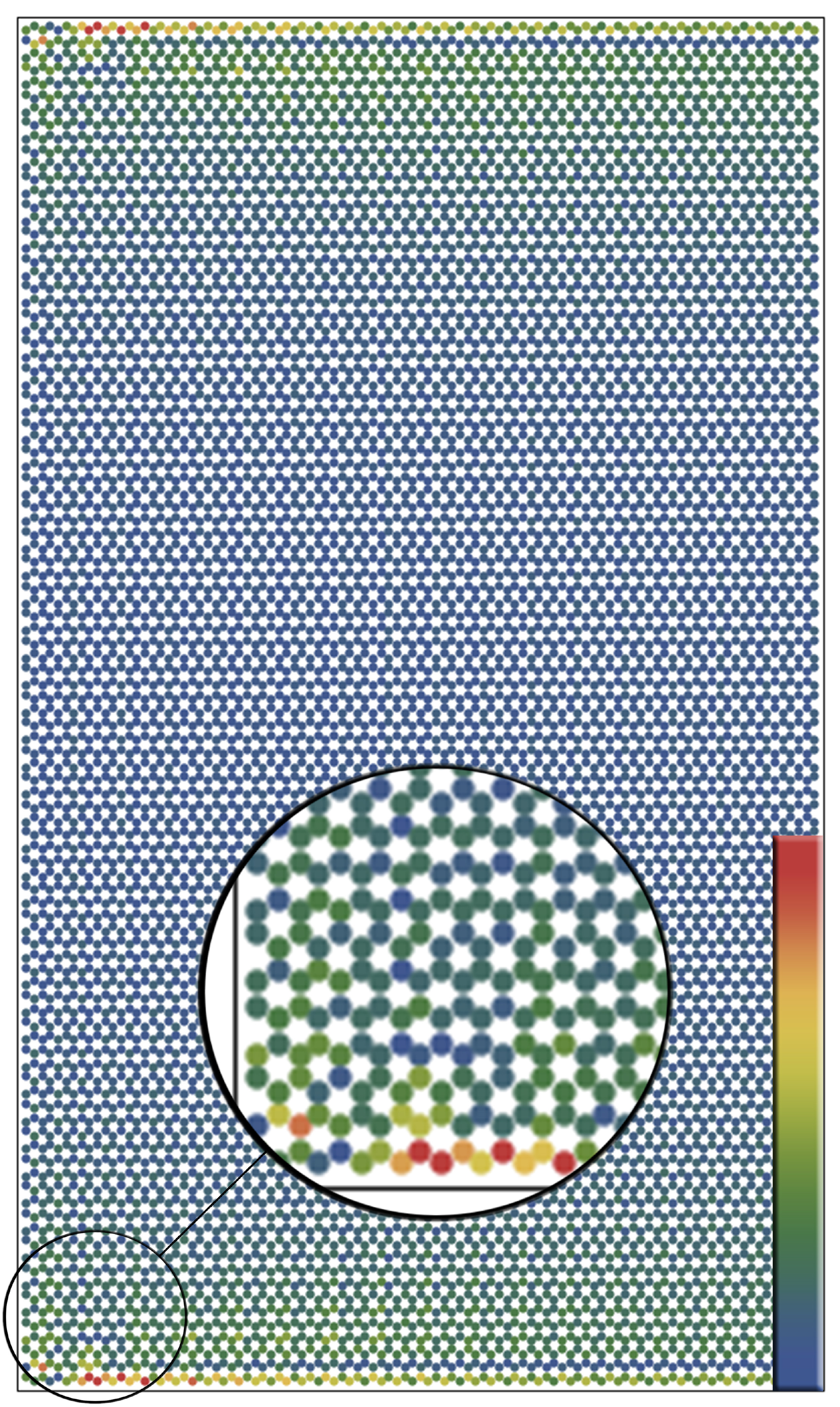}}
\subfloat[]{\label{fig-WF-ZZ-W1}\includegraphics[width=0.25\columnwidth,clip]{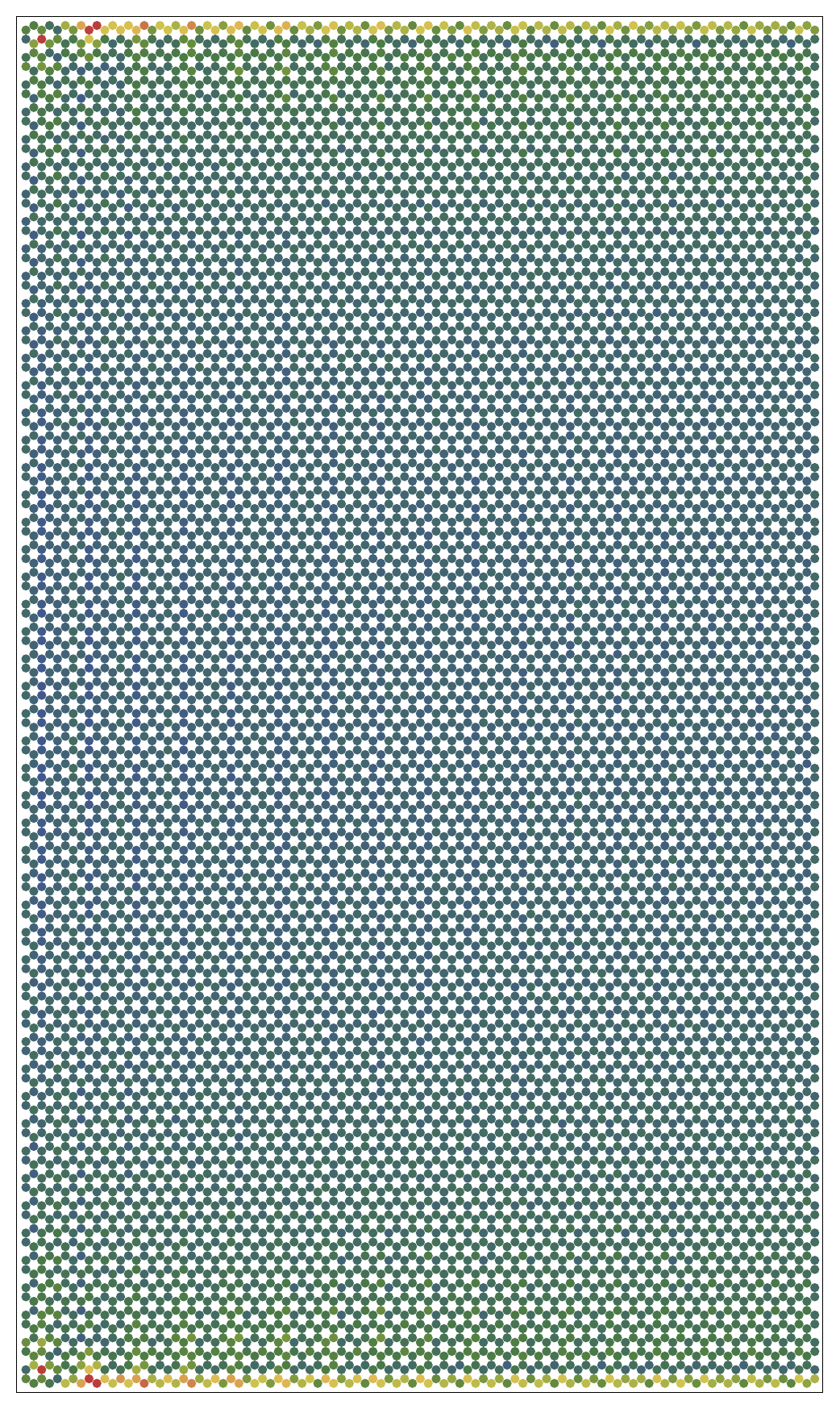}}
\subfloat[]{\label{fig-WF-ZZ-W5}\includegraphics[width=0.25\columnwidth,clip]{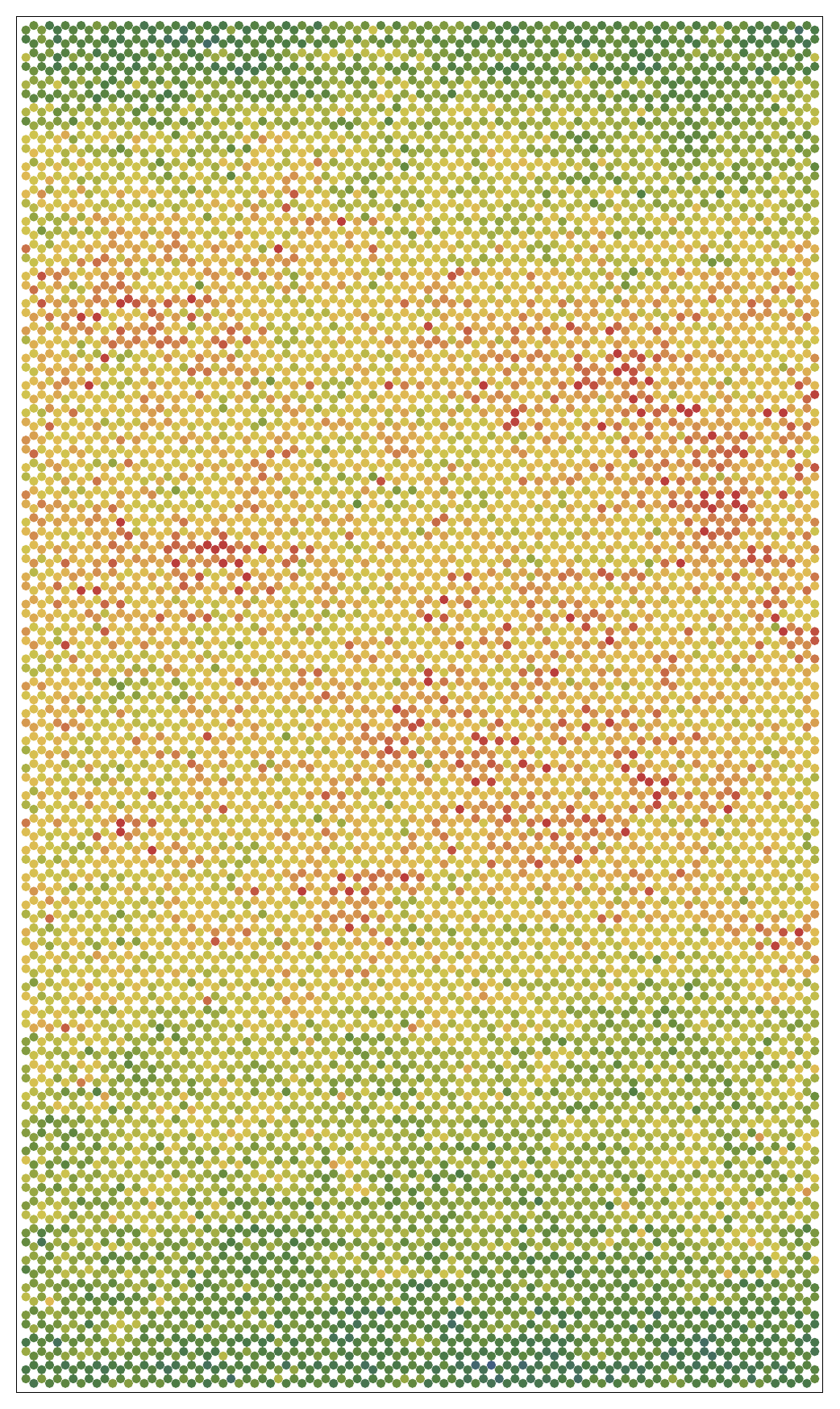}}
\subfloat[]{\label{fig-WF-ZZ-W10}\includegraphics[width=0.25\columnwidth,clip]{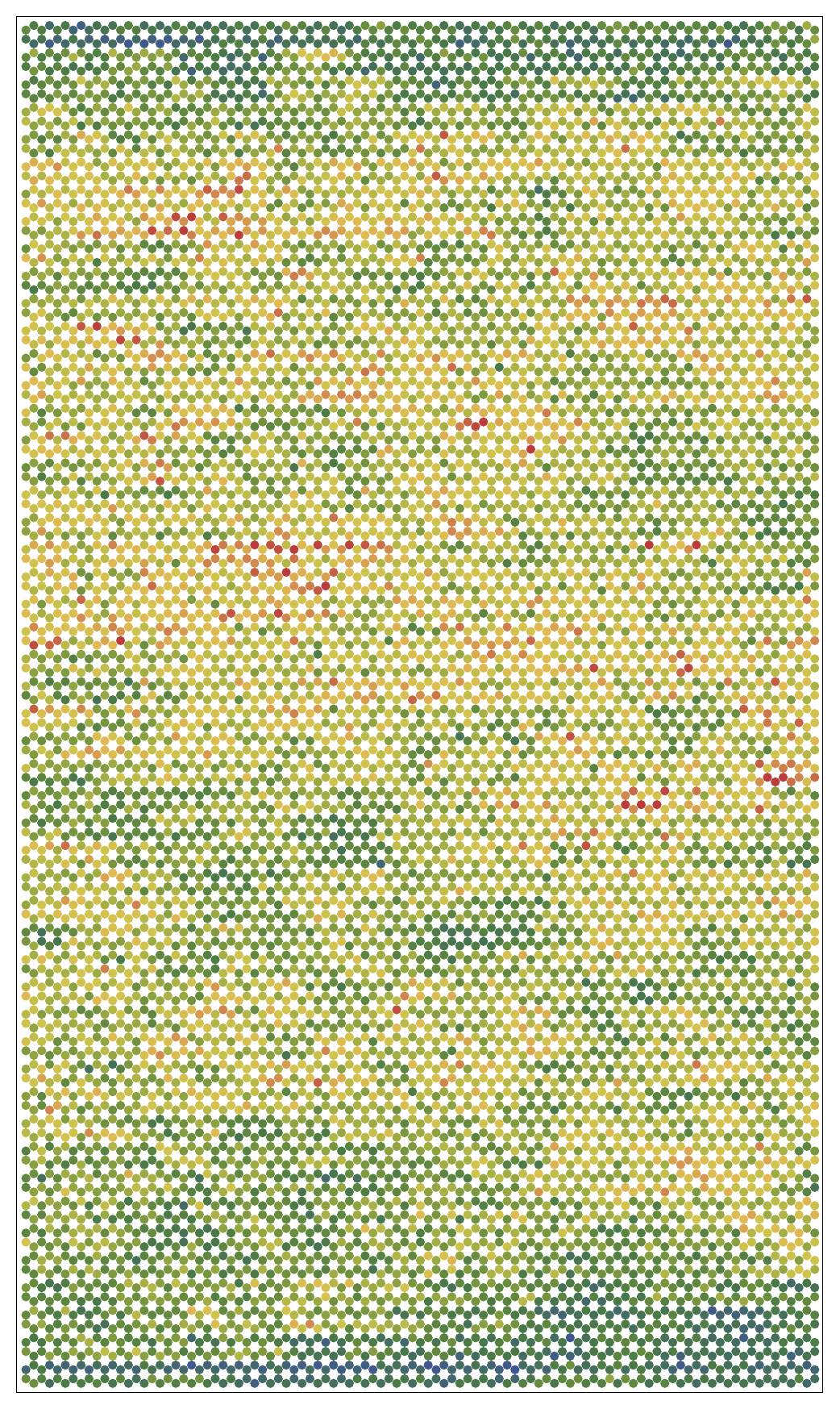}}
\caption{Disorder-averaged $|\psi_{l,m}|^2$ values for $500$ ZZ graphene samples
with $10^4$ lattice sites at $E=0$ and (a)~$W=0.5$, (b)~$W=1$,  (c)~$W=5$ and
(d)~$W=10$. Each wave function has been normalized prior to averaging. The large
circle in panel~(a) shows a zoom of the area in the bottom left corner and the
colour scale on the bottom right of (a) indicates the values of $|\psi|^2$ from
$0$ (blue) to $1$ (red) used for all panels. The transport direction $m$ is
along the horizontal in all panels.}
\label{fig-WF-ZZ}
\end{figure}
%%%%%%%%%%%%%%%%%%%%%%%%%%%%%%%%%%%%%%%%%%%%%%%%%%%%%%%%%%%%%%%%%%%%%%%%%%

Once the modified TMM has reached convergence, the wave functions $(\psi_l,
 \psi_{l-1})$ are true eigenfunctions of the global $2M \times 2M$
forward-backward transfer matrix $\mathcal{T}_{L}^{\dagger} \mathcal{T}_{L}$
for a given sample. Hence $\psi_{l,m}$,
$l,m=1,\ldots, M$, is the
eigenfunction of $\mathcal{H}$.
In fig.~\ref{fig-WF-ZZ} we show the results for ZZ graphene at four different
values of disorder at $E=0$. For weak disorder $W=0.5$ and $1$, one can clearly
see the enduring presence of edge states
previously predicted for clean ZZ samples~\cite{BreF06}. For stronger
values of disorder, the spatial disorder distribution itself becomes dominant. At
$E=0.25$ there is no evidence of edge states.
Results for AC graphene are similarly consistent with the literature, i.e.\ we
find an absence of edge states for the chosen AC graphene lattice sizes
consistent with semiconducting behaviour on finite width samples~\cite{BreF06}.
As expected for square lattices, we do not observe those strong edge states.

%%%%%%%%%%%%%%%%%%%%%%%%%%%%%%%%%%%%%%%%%%%%%%%%%%%%%%%%%%%%%%%%%%%%%%%%%%
\section{Conclusions}
%%%%%%%%%%%%%%%%%%%%%%%%%%%%%%%%%%%%%%%%%%%%%%%%%%%%%%%%%%%%%%%%%%%%%%%%%%

Our results show that up to lengths scales of $1500$ times the C-C distance in
graphene, i.e.\ up to $213\,$nm, onsite disordered graphene, even with
inter-valley scattering, exhibits surprisingly delocalised states in the
vicinity of the Dirac point. This corroborates the trend towards
similar such delocalisation-like behaviour found
previously~\cite{AmiJS09,AmaE09,Hil10,SonSF11,BarN12,AmaKKE13}, while also
reaffirming that the true infinite system limit obeys the localisation
predictions~\cite{XioX07,BarTBB07,NomKR07,LheBNR08,SchSF09,GunW10}.
In fact, the tendency for large localisation lengths is so strong that even FSS
can mislead to construct seemingly extended branches, although a very large
system size analysis shows that only the localised behaviour corresponds to the
true thermodynamic behaviour~\cite{BarTBB07,NomKR07}.
We emphasise that our results also explain graphene's robustness against defects
in similarly sized ribbons~\cite{WanOJW11,LiWZL08}, billiards~\cite{MiaWZC07}
and quantum dots~\cite{PonSKY08}.
Our approach is based on a modified TMM which allows to study "square" flakes
of graphene. This TMM can convincingly reproduce the infinite-size estimates of
localisation lengths obtained from standard TMM and we expect the method to be
useful in other contexts as well.

\begin{acknowledgments}

CGS and RAR are grateful to the CSC and UCM for hospitality, respectively, and
to Ministerio de Educaci\'{o}n, Comunidad de Madrid and the European Social Fund
for funding the research stays at Madrid and Warwick during which much of this
work was done. Work at Madrid was supported by MICINN (project MAT2010-17180).
Calculations were done at (i)~CSC Warwick (MidPlus), (ii)~cluster for physics
(UCM, Feder FUNDS and CEI Moncloa) and (iii)~Centro Nacional de
Supercomputaci\'{o}n - Barcelona Supercomputing Center.

\end{acknowledgments}


\begin{thebibliography}{10}
\bibitem{foot1}
A ``square'' $M\times M$ sample for, e.g.\ ZZ graphene physically
corresponds to a rectangle  with length to width ratio of $\sqrt{3}M/(3M-2)\sim
0.58$. For AC graphene, the ratio is $(3M-2)/\sqrt{3}M$.
\bibitem{foot2}
The choice of  layers in fig.~\ref{fig-lattice} leads to equal spacing
for ZZ graphene  with inter-layer distance $0.142$nm $\times \cos \pi/6 =
0.123$nm. For AC  graphene the inter-layer spacing alternates between $0.142$nm
and $0.142$nm  $\times \sin \pi/6=0.071$nm.  Similar considerations apply in the
transverse direction. We will not  attempt to rescale these length scales here.
\bibitem{foot3}
Previous studies have demonstrated that the finite width of AC flakes
determines its metallic or insulating behaviour at $E=0$ besides the
disorder~\cite{NakFD96}. In the case that $2M_{\text{AC}}=3n-1$ with $n\in
\mathbb{Z}$, the system is metallic. At weak disorder, we observe this effect
in our calculation through a large increase of $\lambda_M$ for such
$M_{\text{AC}}$ that satisfies the above condition. At $E=0.25$ we do not see
these finite width effects. Here we always choose $2M_{\text{AC}}\neq 3n-1$ in
order to avoid the edge-state metallicity at these special sizes for
$W=0$~\cite{NakFD96}.
\bibitem{foot4}
The calculation of a single $M\times L= 700 \times 700$ sample for ZZ
graphene can  take more than $6$ hours at small $W\lesssim 2$ one a single
processor core.  The computing time increases to about a week for some
$1500\times 1500$ samples.
\end{thebibliography}
\end{document}